\documentclass{article}
\usepackage{amsfonts}
\usepackage{amsmath}

\begin{document}

\title{On data analysis and variable selection: the minimum entropy analysis}
\author{Chih-Yuan Tseng$^1$ and CC Chen$^2$ \\
$^1$Department of Physics and \\
$^2$Department of Earth Sciences and Graduate Institute of Geophysics, \\
National Central University, Jhongli 320, Taiwan, ROC}
\date{}

\maketitle

\begin{abstract}
In this work, we present a minimum entropy analysis scheme for variable
selection and preliminary data analysis. The variable selection can be
achieved by the increasing preference of variables. We show such a
preference to has a unqiue form, which is given by the entropy of models
associated with variables. Evaluating the entropy provides a complete
ranking scheme of variables. This scheme not only indicates preferred
variables but also may reveal the system's nature and properties. We
illustrate the proposed scheme to analyze a set of geological data for three
carbonate rock units in Texas and Oklahoma, and compare to the discriminant
function analysis. The result suggests this scheme to provide a quick 
and 
robust analysis, and the use in data analysis is promising.

\end{abstract}

\section{Introduction}

When one investigates an unknown system according to experimental
observations made on this system, two questions are commonly addressed. How
does one model the system? The model associates the experimental
observations with corresponding experimental responses. Unfortunately, there
is no systematical method to answer it. It is usually resolved through the
methods of trials and errors, empirical regressions, and some intuitive
assumptions etc. The second question is that suppose a model is found, which
experimental observations shall be codified into the model. This is exactly
a variable selection question. Here we will put the first question aside,
and focus on resolving the second question.

A problem similar to variable selection has been tackled through different
approaches in the past. It is the model selection problem. Several methods
such P-value, Bayesian approach, and Kullback-Leibler distance based
approach et al. are some examples. The P-values method selects model by
comparing probability of model given a null model and experimental data sets
to a threshold value assessed from same data sets \cite{Raftery95}. Yet
since this method is restricted to two models and required ad hoc rules to
assess threshold value, people has developed the Bayesian approaches to
overcome these defects (\cite{Raftery95} and \cite{Bayesian}). The Bayesian
method applies Bayes theorem to update our beliefs and uncertainty about
models from prior distributions generated from some prior modeling rules
first. A preferable model, thereafter, is chosen according to Bayes factor,
ratio of posterior distributions of different models. Bayesian Information
Criterion (BIC) is one of most popular Bayesian based model selection
criteria (\cite{Raftery95}, Kieseppa, Forbes and Peyrard's works in \cite%
{Bayesian}). Yet all of these methods require prior information generated
from some ad hoc prior modeling rules that suits people's need.

Aside from Bayesian framework, people has developed relative entropy, mutual
information, or Kullback-Leibler distance based approach (\cite{Bonnlander94}%
, \cite{Dupuis03}, and An Information Criterion (AIC) of Akaike \cite%
{Akaike74}). The rationale is to design a criterion based on aspect of
entropy. The reason of employing relative entropy for selection criterion is
discussed in \cite{Tseng06}. Recently, another criterion for model
selection, CIC, is proposed by Rodriguez based on aspect of information
geometry \cite{Rodriguez05}. That is a generalized version of AIC and BIC.
Preferred model is selected to minimize a quantity derived from Bayes's
theorem.

In the case of variable selection, Dupuis and Robert \cite{Dupuis03}
proposed to model the system with different combinations of experimental
observations, the variables selected from a set of variables. It generates
several models associated with different combinations of variables. Thus
variable selection problem becomes the problem of model selection.
Evaluating the Kullback-Leibler distance between full model described by a
complete set of variables for interested system and it's approximations,
submodels, described by subset of variables. When the full model is
tractable, the preferred submodel is selected when it's Kullback-Leibler
distance reaches a threshold value. This threshold value is usually
estimated by experiences from experimental data. Since submodels are
projections of full model, there is no need the prior modeling rule to
generate prior distribution for submodel. Yet one still requires prior
information on full model. In addition, when there is no complete set of
variables, namely, only limited experimental measurements regarding to the
system can be conducted, this strategy becomes inadequately.

Our goal is to apply the method of maximum entropy (ME) to design a tool for
variable selection that is free from defects in Dupuis and Robert's
approach. Following the axiomatic approach proposed in developing method of
ME\ to a tool for assigning a probability to a system \cite{Jaynes57} and a
tool for updating probability \cite{ME}, Tseng has showed an entropic
criterion for model selection \cite{Tseng06}. Based on this study, we
generalize it to obtain a Minimum Entropy Analysis (MEA) in Sec.2. The
proposed analysis tool provides a complete ranking scheme of variables. It
not only allows to select preferred variables but also to suggest an
analysis scheme to reveal nature and properties of the system. To illustrate
the MEA scheme, we will study a binary system in the geology in Sec.3. Some
discussions are made afterward. Our demonstration shows the MEA scheme to be
a promising tool in data analysis. Finally some conclusions are given.

\section{The minimum entropy analysis}

\subsection{Basic features}

Despite designing a pertinent criterion for selection, the selection also
can be achieved by the increasing preference of variables. Since properties
and meaning of the variables, experimental observations, are sometimes quite
different, it is meaningless to compare variables directly. For example,
suppose two experimental observations, mass and area, are measured for
studying a system. How does one evaluate weightings of these two quantities
to determine the dominant quantity in the model given to study this system?
Namely, what is the preference of these variables? In Dupuis and Robert's on
variable selection problem \cite{Dupuis03}, they proposed codifying the
variables by a specific model such as logit model for a linear binary
system. Afterward, ranking variables by evaluating the Kullback-Leibler
distance between the model described by complete set of variables and it's
projections, submodels, described by subset of variables. Yet when there is
no complete set of variables, namely, experimental measurements only provide
limited information regarding to the system, this strategy becomes
inadequately.

The approach for model selection proposed by Tseng may spells out a
resolution in variable selection problem \cite{Tseng06}. It states that
suppose a family of probability models is found to interpret the system $%
\left\{ P^{m}\left( x\right) \right\} $, where m labels the model and x
denotes states of the model. The preference of models is uniquely
determined, which is in the form of relative entropy of model $P^{m}\left(
x\right) $ and a uniform reference measure $\mu $,

\begin{equation}
S\left[ P^{m}\left\vert \mu \right. \right] =-\sum_{x}P^{m}\left( x\right)
\ln \frac{P^{m}\left( x\right) }{\mu }\text{ .}  \label{S[P|u]}
\end{equation}
This scalar value measures differences between models and a uniform
reference measure. Maximizing the relative entropy indicates $P^{m}\left(
x\right) $ to equal to $\mu $. Namely, there is no information regarding to
the system being codified into $P^{m}\left( x\right) $. Within the family,
when the relative entropy is decreased, there is more information of the
system being codified into $P^{m}\left( x\right) $ One can then rank those
probability models according to decreasing $S\left[ P^{m}\left\vert \mu
\right. \right] $ value.

\subsection{Logic behind the MEA}

Based on Dupuis and Robert's approach, one can determine the preference of
variables by first codifying these variables into a model. This model can be
any function that optimally associates the experimental observations,
variables, and responses. According to Tseng's work on model selection \cite%
{Tseng06}, one needs to further codify this model into a probability
distribution of observing the experimental responses given the variables in
order to compute the preference. Thus the preference of variables is
determined form the preference of those probability distributions.

Based on these aspects, the logic behind the proposed MEA scheme then
involves two stages. The first stage is to determine a probability model
that associates experimental observations and responses. The method of ME
proposed by Jaynes \cite{Jaynes57} provides a solution. Since the method of
ME\ requires the information that will be codified into the probability
distribution to be in the form of constraint, it turns the question of
probability assignment into a question of what is constraint. We will come
back this point later.

Next we follow the axiomatic approach \cite{ME} to determine the form of
preference of the probability distributions. The basic strategy (Skilling of 
\cite{ME}) is one of induction: (1) if a general rule exists, then it must
apply to special cases; (2) if in a certain special case we know which is
the best approximation, then this knowledge can be used to constrain the
form of preference; and finally, (3) if enough special cases are known, then
preference will be completely determined. The known special cases are called
the \textquotedblleft axioms\textquotedblright\ of ME. The axioms used here
must reflect the conviction that one should not change one's mind
frivolously, that whatever information was originally codified into the
exact probability distribution is important and should be preserved. As
shown in \cite{Tseng06}, three axioms are employed: (1) local information
has local effects; (2) the ranking should not depend on the coordinates of
the system, and (3) consistency for independent subsystems. The functional
form for the preference is therefore uniquely determined, which has the form
of relative entropy, Eq.(\ref{S[P|u]}). Please refer to Caticha of \cite{ME}
for detail proof on the axiomatic approach.

\subsection{The MEA scheme}

Suppose a model, function of $f\left( \mathbf{\hat{x}},\hat{\beta}\right) $,
is given to associate N experimental observations denoted by variable $%
\mathbf{\hat{x}}=\left\{ \mathbf{x}_{1},\mathbf{x}_{2},\cdots \mathbf{x}%
_{N}\right\} $ and parameters $\hat{\beta}=\left\{ \beta _{1},\beta
_{2},\cdots \beta _{N}\right\} $ with experimental responses $\mathbf{M}%
\left( \mathbf{\hat{x}},\hat{\beta}\right) $. Each observations is repeated $%
l$ times, which give $l$ measurements $\mathbf{x}_{i}=\left\{
x_{i}^{1},x_{i}^{2},\cdots x_{i}^{l}\right\} $ and corresponding responses $%
\mathbf{M}\left( \mathbf{\hat{x}},\hat{\beta}\right) =\left\{
M^{1},M^{2},\cdots M^{l}\right\} $. The response \textbf{M} will be either
\textquotedblleft 1\textquotedblright\ for positive response or
\textquotedblleft 0\textquotedblright\ for negative response. Notes that one
way to determine $\hat{\beta}$ is through method of Maximum Likelihood
Estimate (MLE) \cite{Johnson99}. For example, the logit model 
\begin{equation}
f_{\text{logit}}\left( \mathbf{\hat{x}},\hat{\beta}\right) =\frac{\exp
\sum_{i=1}^{N}\beta _{i}\mathbf{x}_{i}}{\exp \sum_{i=1}^{N}\beta _{i}\mathbf{%
x}_{i}+1}\text{ ,}  \label{Logit model}
\end{equation}%
is usually given as a regression model for a binary output system. Given
these N variables, there will be $2^{N}-2$ different combinations (subsets)
of variables $\mathbf{\hat{x}}_{s_{i}}\in \mathbf{\hat{x}}$ to be chosen
from $\mathbf{\hat{x}}$, and form $2^{N}-2$ submodels $f\left( \mathbf{\hat{x%
}}_{s_{i}},\hat{\beta _{s_{i}}}\right) $.

Afterward we have to determine the probability $P_{s}\left( \mathbf{\hat{x}}%
_{s_{i}}\hat{\left\vert \beta _{s_{i}}\right. }\right) $ of observing the
experimental responses $\mathbf{M}$ given the subset of variables $\mathbf{%
\hat{x}}_{s_{i}}$. In the framework of method ME, one has to identify the
relevant information to be codified into the probability distribution. In
this example, it is obvious that the experimental responses are the quantity
we need to know about the system, that can be written in the form of
constraint

\begin{equation}
\left\langle \mathbf{M}\right\rangle =\sum f\left( \mathbf{\hat{x}}_{s_{i}},%
\hat{\beta _{s_{i}}}\right) P_{s}\left( \mathbf{\hat{x}}_{s_{i}}\hat{%
\left\vert \beta _{s_{i}}\right. }\right) \text{ ,}
\end{equation}%
where $\left\langle \mathbf{M}\right\rangle $ is the expectation value of
the responses. Thus the method of ME\ indicates the preferred $P_{s}\left( 
\mathbf{\hat{x}}_{s_{i}}\hat{\left\vert \beta _{s_{i}}\right. }\right) $ to
be

\begin{equation}
P_{s}\left( \mathbf{\hat{x}}_{s_{i}}\hat{\left\vert \beta _{s_{i}}\right. }%
\right) =\frac{\exp -\lambda f\left( \mathbf{\hat{x}}_{s_{i}},\hat{\beta
_{s_{i}}}\right) }{Z}\text{ ,}  \label{Psubmodel-ME}
\end{equation}%
where partition function $Z=\sum_{\hat{x_{s_{i}}}\in \hat{x}}\exp -\lambda
f\left( \mathbf{\hat{x}}_{s_{i}},\hat{\beta _{s_{i}}}\right) $ and $\lambda $
is a Largrangian multiplier, which is set to one for sake of simplicity.
Alternatively, the probability distribution can be obtained by simply
normalizing submodels $f\left( \mathbf{\hat{x}}_{s_{i}},\hat{\beta _{s_{i}}}%
\right) $, 
\begin{equation}
P_{s}^{\prime }\left( \mathbf{\hat{x}}_{s_{i}}\left\vert \beta
_{s_{i}}\right. \right) =\frac{f\left( \mathbf{\hat{x}}_{s_{i}},\hat{\beta
_{s_{i}}}\right) }{Z^{\prime }}\text{ ,}  \label{Psubmodel-approx}
\end{equation}%
where normalization constant $Z^{\prime }=\sum_{\hat{x_{s_{i}}}\in \hat{x}%
}f\left( \mathbf{\hat{x}}_{s_{i}},\hat{\beta _{s_{i}}}\right) $ as well.
Notes that this form of probability distribution is actually just a first
order approximation of Eq.(\ref{Psubmodel-ME}).

Thus the increasing ranking order of these submodel $f\left( \mathbf{\hat{x}}%
_{s_{i}},\hat{\beta _{s_{i}}}\right) $ is given by the decreasing relative
entropy Eq.(\ref{S[P|u]}) with $P^{m}$ being replaced by $P_{s}\left( 
\mathbf{\hat{x}}_{s_{i}}\left\vert \beta _{s_{i}}\right. \right) $ or $%
P_{s}^{\prime }\left( \mathbf{\hat{x}}_{s_{i}}\left\vert \beta
_{s_{i}}\right. \right) $. Furthermore, one can easily rewrite Eq.(\ref%
{S[P|u]}) into 
\begin{equation}
S\left[ P_{s}\left\vert \mu \right. \right] =S\left[ P_{s}\right] +\ln \mu 
\text{ ,}  \label{S[Ps|Puni]=S[Ps]}
\end{equation}%
where $S\left[ P_{s}\right] =-\sum_{\mathbf{\hat{x}}_{s_{i}}\in \mathbf{\hat{%
x}}}P_{s}\left( \mathbf{\hat{x}}_{s_{i}}\left\vert \beta _{s_{i}}\right.
\right) \ln P_{s}\left( \mathbf{\hat{x}}_{s_{i}}\left\vert \beta
_{s_{i}}\right. \right) $ entropy of the submodel. Because $\ln \mu $ is a
constant value for a uniform reference measure, the preference of submodels
is equivalent to the decreasing $S\left[ P_{s}\right] $ value.

Evaluating the entropy of all submodels, a complete ranking scheme of
different subsets of variables $\mathbf{\hat{x}}_{s_{i}}$ is determined.
Preferred subset of variables then can be identified that is the one that
has minimum entropy value within this set of variables. Notice that the use
of this scheme is not totally exhausted yet. By further analyzing the
ranking scheme of different subsets $\mathbf{\hat{x}}_{s_{i}}$, one may
determine significance of different combinations of variables that are
codified into the submodel. The nature and properties of the system may
thereafter be revealed through this analysis. For example, as we know,
correlation functions between two variables arbitrarily chosen within $%
\mathbf{\hat{x}}$ may reveal some properties of the system. Although we did
not compute the correlation functions in this scheme, the preference of
different correlations is still implicitly spelled out by the ranking scheme
of different subsets. One can attribute this to when the model is given to
associate the variables and responses, the correlations between the
variables are defined in the model. Thus determining the ranking scheme of
different combinations of variables indirectly indicate the significance of
different correlations.

Thus one may treat the MEA\ scheme as a quick data analysis tool. It
provides preliminary information about the system. This use is implicitly in
some other approaches mentioned previously. We will illustrate the use of
the MEA\ scheme in detail by studying a geological problem next.

\section{A special case: a binary system in Geology}

\subsection{The problem}

Considering a geological example of sample classification (Davis of \cite%
{Geology}). We briefly address the result by means of a standard tool, the
Discriminant Function Analysis (DFA), for classifying geological samples.
Then we show how to extract important variables in determining the category
of samples using our minimum entropy analysis. Furthermore, some information
regarding the formation of sample rocks. Comparing both results from the DFA
and the MEA improves our understanding and enhances our confidence in our
MEA scheme.

Saltwater is trapped in sedimentary rocks at the time they are formed in the
marine environment. The chemical composition of the connate water is
subsequently modified by ion exchange and other reactions, by mixing with
other brines, and by dilution by infiltrating surface waters. Brines
recovered during drillstem tests of wells may have relict compositional
characteristics that provide clues to the origin or depositional environment
of their source rocks. Table 1 contains brine analyses for oil-field waters
from three groups of carbonate units in Texas and Oklahoma (Davis of \cite%
{Geology}). The first column in Table 1 denotes the brine samples belonging
or not belonging to some specific carbonate unit, Unit G.

\subsection{The discriminant function analysis and results}

The discriminant function analysis combines a rationale similar to that of
analysis of data variance with computational procedures based on eigenvector
calculations, e.g. the PCA (principle component analysis). Multivariate
measurements made on the samples alone, such as the brine data in Table.\ref%
{carbonate}, can be used in the DFA to find combinations of measurements
that allow the various categories of samples to be distinguished. The
problem of DFA is basically one of finding a set of linear weights for the
variables that causes a multivariate analogue of the F-ratio to be a
maximum. A succession of discriminant functions along which the samples are
as distinct as possible, can be thus calculated and each represents
successively the most efficient discriminator possible. For many calculation
details, please refer to the book of Davis \cite{Geology}.

The DFA can be applied to those data in Table.\ref{carbonate} to determine
if they are distinctive. The first discriminant function thus calculated is
an inner product of (-0.3765, -0.0468, 0.0112, -0.0148, -0.0174, -0.0110)$%
\cdot$ $(\text{HCO}_{3}, \text{SO}_{4}, \text{Cl}, \text{Ca}, \text{Mg}, 
\text{Na})^{T}$, which can clearly separates samples from Unit G and other
units. Please note that the weighting factors in the first discriminant
function for variables HCO$_{3}$ and SO$_{4}$, i.e. -0.3765 and -0.0468,
represent the first two largest factors in magnitude among six, thus
indicating those two variables play the most dominant effect in
classification.

\vspace{-8pt} 
\begin{table}[h]
\caption{\baselineskip=8pt{\protect\footnotesize \textsf{Chemical analyses
of brines (in ppm) recovered from drillstem tests of three carbonate rock
units in Texas and Oklahoma. Adapted from Davis of \protect\cite{Geology}. }}
}
\label{carbonate}
\centering
\begin{tabular}{|c|c|c|c|c|c|c|}
\hline
Unit G & HCO$_{3}$ & SO$_{4}$ & Cl & Ca & Mg & Na \\ \hline
N & 10.4 & 30 & 967.1 & 95.9 & 53.7 & 857.7 \\ \hline
N & 6.2 & 29.6 & 1174.9 & 111.7 & 43.9 & 1054.7 \\ \hline
N & 2.1 & 11.4 & 2387.1 & 348.3 & 119.3 & 1932.4 \\ \hline
N & 8.5 & 22.5 & 2186.1 & 339.6 & 73.6 & 1803.4 \\ \hline
N & 6.7 & 32.8 & 2015.5 & 287.6 & 75.1 & 1691.8 \\ \hline
N & 3.8 & 18.9 & 2175.8 & 340.4 & 63.8 & 1793.9 \\ \hline
N & 1.5 & 16.5 & 2367 & 412 & 95.8 & 1872.5 \\ \hline
Y & 25.6 & 0 & 134.7 & 12.7 & 7.1 & 134.7 \\ \hline
Y & 12 & 104.6 & 3163.8 & 95.6 & 90.1 & 3093.9 \\ \hline
Y & 9 & 104 & 1342.6 & 104.9 & 160.2 & 1190.1 \\ \hline
Y & 13.7 & 103.3 & 2151.6 & 103.7 & 70 & 2054.6 \\ \hline
Y & 16.6 & 92.3 & 905.1 & 91.5 & 50.9 & 871.4 \\ \hline
Y & 14.1 & 80.1 & 554.8 & 118.9 & 62.3 & 472.4 \\ \hline
N & 1.3 & 10.4 & 3399.5 & 532.3 & 235.6 & 2642.5 \\ \hline
N & 3.6 & 5.2 & 974.5 & 147.5 & 69 & 768.1 \\ \hline
N & 0.8 & 9.8 & 1430.2 & 295.7 & 118.4 & 1027.1 \\ \hline
N & 1.8 & 25.6 & 183.2 & 35.4 & 13.5 & 161.5 \\ \hline
N & 8.8 & 3.4 & 289.9 & 32.8 & 22.4 & 225.2 \\ \hline
N & 6.3 & 16.7 & 360.9 & 41.9 & 24 & 318.1 \\ \hline
\end{tabular}%
\end{table}

\subsection{The minimum entropy analysis}

According to conventional studies (\cite{Dupuis03} and \cite{Johnson99}),
one can pertinently associate six experimental observations with binary
responses in this geological example through the logit model, Eq.(\ref{Logit
model}). The variables $\mathbf{\hat{x}}_{G}\mathbf{=}\left\{ \mathbf{x}_{1},%
\mathbf{x}_{2},\cdots \mathbf{x}_{6}\right\} $ denotes observations on
contents of six chemical compounds $\left\{ \text{HCO}_{\text{3}},\text{SO}_{%
\text{4}},\text{Cl},\text{Ca},\text{Mg},\text{Na}\right\} $ correspondingly.
As shown in Table.\ref{carbonate}, nineteen measurements were made. There
are five positive experimental responses, denoted by symbol
\textquotedblleft Y\textquotedblright\ and fourteen negative responses. The
method of ME\ suggests that the preferred probability distribution of
observing the experimental responses given variables $\mathbf{\hat{x}_{G}}$
to give

\begin{equation}
P\left( \mathbf{\hat{x}}_{G}\hat{\left\vert \beta \right. }\right) =\frac{%
\exp -f_{\text{logit}}\left( \mathbf{\hat{x}}_{G},\hat{\beta}\right) }{Z}%
\text{ ,}  \label{PsG_ME}
\end{equation}%
where partition function $Z=\sum_{\mathbf{\hat{x}}_{G}}\exp -f_{\text{logit}%
}\left( \mathbf{\hat{x}}_{G},\hat{\beta}\right) $. Or normalizing the logit
model within this data set gives

\begin{equation}
P^{\prime }\left( \mathbf{\hat{x}}_{G}\hat{\left\vert \beta \right. }\right)
=\frac{f_{\text{logit}}\left( \mathbf{\hat{x}}_{G},\hat{\beta}\right) }{%
Z^{\prime }}=\frac{1}{Z^{\prime }}\frac{\exp \sum_{i=1}^{6}\beta _{i}\mathbf{%
x}_{i}}{\exp \sum_{i=1}^{6}\beta _{i}\mathbf{x}_{i}+1}\text{ ,}
\label{PsG_approx}
\end{equation}%
where normalization constant%
\begin{equation}
Z^{\prime }=\sum_{\mathbf{\hat{x}}}\frac{\exp \sum_{i=1}^{6}\beta _{i}%
\mathbf{x}_{i}}{\exp \sum_{i=1}^{6}\beta _{i}\mathbf{x}_{i}+1}\text{ }.
\end{equation}%
Given these six variables, $2^{6}-2=62$ different combinations of variables $%
\mathbf{\hat{x}}_{s_{i}}^{G}\in \mathbf{\hat{x}}_{G}$ are obtained.
Thereafter, one can generate 62 probability submodels $P_{s_{i}}\left( 
\mathbf{\hat{x}}_{s_{i}}^{G}\right) =P\left( \mathbf{\hat{x}}_{s_{i}}^{G}%
\hat{\left\vert \beta \right. }\right) $ or $P^{\prime }\left( \mathbf{\hat{x%
}}_{s_{i}}^{G}\hat{\left\vert \beta \right. }\right) $.

Evaluating the entropy of $P_{s_{i}}\left( \mathbf{\hat{x}}%
_{s_{i}}^{G}\right) $, Eq.(\ref{S[Ps|Puni]=S[Ps]}), with different subsets
of variables $\mathbf{\hat{x}}_{s_{i}}^{G}$ gives ranking order of different
submodels $P_{s_{i}}\left( \mathbf{\hat{x}}_{s_{i}}^{G}\right) $. The
coefficient $\beta _{i}$ are determined through fitting the logit model to
experimental measurements by MLE method (in Appendix: software for ordinal
data modeling of \cite{Johnson99} - a MATLAB function of \textquotedblleft
Maximum likelihood estimation and model criticism\textquotedblright ). The
result is listed in Table.\ref{Carbonate ranking}. Here we only list 18 out
of 62 submodels. We calculate the entropy of two probability distributions,
Eq.(\ref{PsG_ME}) and (\ref{PsG_approx}), in second and third column
respectively. The ranking scheme is in the order of decreasing entropy
value. To analyze this ranking scheme, we proceed with a two-step approach.
In first step, we examine the submodel that has the minimum entropy value.
In this example, there are 16 out of 62 submodels that has the minimum
entropy value, 2.866 or 1.791 in the second and third column of Table.\ref%
{Carbonate ranking} respectively. Notes that since the minimum significant
figure of experimental data in Table.\ref{carbonate} is three, the entropy
value should also has three significant figures and forth digit is just an
estimate. The preference of these 16 submodels are indistinguishable. Notes
that the digits in round bracket shows a numerical result when the
significant figure is not considered. It just indicates that if the
significant figures are higher the resolution of entropy will be better.
Thus the preference of these 16 submodel then still can be identified. That
will further aid the analysis of preference in detail.

In order to determine the most dominant variables from these 16 submodels,
frequencies of six variables appeared in these 16 submodels are recorded in
the present. The frequencies for observing first and second variable are 16
and 15 respectively and 8 for rest of variables. This result suggests that
the ability of interpreting the experimental measurements by the logit model
is strongly dominated by the first variable, HCO$_{\text{3}}$, and the
second variable, SO$_{\text{4}}$. The variables 3 to 6 seem to play a minor
role here. This is exactly the result obtained though the DFA\ analysis
mentioned previously. Yet the MEA\ scheme is more straightforward.

In second step, we analyze the ranking scheme further to identify preference
of first and second variable. Since the variables 3 to 6 play a minor role
here, we concentrate on first two variables. We list two more submodels that
only include the first and second variable, HCO$_{\text{3}}$ and SO$_{\text{4%
}}$, respectively in Table.\ref{Carbonate ranking}. The entropy value in
third column shows a dramatic changes from 2.328 of submodel \{010000\} that
only has second variable, 2.060 of \{100000\} that has only first variable,
to 1.791 of \{110000\} for the case of two variables being simultaneously
included. The same trend is also observed in the second column although no
dramatic changes is observed. The ranking scheme indicates that the first
variable HCO$_{\text{3}}$ should play a more important role than the second
variable SO$_{\text{4}}$ in the model.

\vspace{-8pt} 
\begin{table}[h]
\caption{\baselineskip=8pt{\protect\footnotesize \textsf{The ranking scheme
of six chemical compounds. First column represent the six chemical
compounds. The number ``1'' denotes the corresponding variables in first row
to be considered and ``0'' denotes to be negelected. Second column present
the entropy value, Eq.(\protect\ref{S[Ps|Puni]=S[Ps]}) of the probability
distribution $P_{s}\left( \mathbf{\hat{x}}_{s_{i}}\right)$ given by Eq.(%
\protect\ref{PsG_ME}) and the third column is entropy value of $%
P^{\prime}_{s}\left( \mathbf{\hat{x}}_{s_{i}}\right)$ given by Eq.(\protect
\ref{PsG_approx}). Each row represents a submodel. Only 18 submodels are
listed.}}}
\label{Carbonate ranking}
\centering%
\begin{tabular}{|cccccc|c|c|}
\hline
HCO$_{\text{3}}$ & SO$_{\text{4}}$ & Cl & Ca & Mg & Na & $S[P_{s}]$ & $%
S[P^{\prime}_{s}]$ \\ \hline
0 & 1 & 0 & 0 & 0 & 0 & 2.893(229075) & 2.328(713745) \\ \hline
1 & 0 & 0 & 0 & 0 & 0 & 2.881(069331) & 2.060(483312) \\ \hline
1 & 1 & 0 & 0 & 0 & 0 & 2.866(92132) & 1.791(857378) \\ \hline
1 & 1 & 0 & 0 & 0 & 1 & 2.866(921309) & 1.791(85668) \\ \hline
1 & 1 & 1 & 0 & 0 & 0 & 2.866(921264) & 1.791(854715) \\ \hline
1 & 1 & 0 & 1 & 0 & 0 & 2.866(921168) & 1.791(849677) \\ \hline
1 & 0 & 1 & 1 & 1 & 1 & 2.866(921139) & 1.791(848879) \\ \hline
1 & 1 & 1 & 1 & 0 & 0 & 2.866(921109) & 1.791(84701) \\ \hline
1 & 1 & 0 & 1 & 0 & 1 & 2.866(921101) & 1.791(846531) \\ \hline
1 & 1 & 0 & 0 & 1 & 0 & 2.866(921112) & 1.791(843955) \\ \hline
1 & 1 & 1 & 0 & 0 & 1 & 2.866(921028) & 1.791(842415) \\ \hline
1 & 1 & 1 & 0 & 1 & 0 & 2.866(921084) & 1.791(842147) \\ \hline
1 & 1 & 0 & 0 & 1 & 1 & 2.866(921072) & 1.791(841769) \\ \hline
1 & 1 & 0 & 1 & 1 & 0 & 2.866(921005) & 1.791(840749) \\ \hline
1 & 1 & 1 & 1 & 1 & 0 & 2.866(920963) & 1.791(83836) \\ \hline
1 & 1 & 0 & 1 & 1 & 1 & 2.866(920964) & 1.791(838358) \\ \hline
1 & 1 & 1 & 1 & 0 & 1 & 2.866(920971) & 1.791(838291) \\ \hline
1 & 1 & 1 & 0 & 1 & 1 & 2.866(920965) & 1.791(838227) \\ \hline
\end{tabular}%
\end{table}

\section{Discussions}

Lithologically speaking, unit G is mainly composed of dolomite (CaMg(CO$_{%
\text{3}}$)$_{\text{2}}$) and anhydrite (CaSO$_{\text{4}}$). In ancient
geological times, Unit G, which is in geology called the \textquotedblleft
Grayburg Dolomite\textquotedblright\ \cite{Geology}, experienced two
important sedimentary processes of dolomitization, which is associated with
the dissolution of calcite by acidic fluids, and evaporation. Anhydrite is
one of the index products from evaporation. Chalcraft and Ward further
claimed that the principal diagenetic processes include dolomitization,
anhydrite occlusion of primary porosity, and leaching \cite{Chalcraft88}.
The dolomitzation plays a crucial role in the formation of unit G, and
followed by anhydrite occlusion of primary porosity. One therefore can infer
that the process of the dissolution of calcite by acidic fluids is more
significant than anhydrite occlusion in the formation of unit G. Namely, the
process involves with chemical compound HCO$_{\text{3}}$ is the most
important one among the six compounds.

In analyzing a set of geological data to seek out the origin or depositional
environment of their source rocks, the MEA suggests that HCO$_{\text{3}}$
and SO$_{\text{4}}$ are two key variables in the occurrence process of Unit
G. The MEA also suggests that HCO$_{\text{3}}$ plays a more important role
than SO$_{\text{4}}$. Therefore, we can conclude that formation of Unit G
may strongly involve with chemical process associated with HCO$_{\text{3}}$.
The chemical process associated with SO$_{\text{4}}$ may then plays a minor
factor in the formation. It is the exact result inferred previously but the
MEA analysis is more straightforward. Similarly, one may conduct further
analysis to extract more information, yet it is out of our scope here.

\section{Conclusions}

The minimum entropy analysis scheme is proposed to analyze experimental data
for extracting information of the corresponding system such as which
experimental observations to play a more important role etc. This is a
question of variable selection, and can be resolved by determining the
preference of these observations. To determine the preference, one
associates the experimental responses and those observations by a
probability model first. Thereafter, as shown in the context, the form of
preference is uniquely determined through the axiomatic approach \cite{ME}.
It is in the form of entropy of probability of observing the experimental
responses given variables. The preferred variables are the one that have
minimum entropy value. Furthermore, since the minimum entropy analysis
present a complete ranking scheme of different combinations of experimental
observations, it indirectly indicates significance of different combinations
of variables in the model. This ranking scheme not only suggests the
preferred variables that should be codified into the model but also may
spell out a route to study the system. Besides, this design resolves two
defects in Dupuis and Robert's approach \cite{Dupuis03} mentioned previously.

We have illustrated the use of the minimum entropy analysis by analyzing a
set of geological data for three carbonate rock units in Texas and Oklahoma.
The MEA scheme indicates the preferred variables most relevant to the
formation of unit G or Grayburg Dolomite to be HCO$_{\text{3}}$ and SO$_{%
\text{4}}$. This result agrees with the result from another well known
analysis tool, the discriminant function analysis. Furthermore, since the
MEA presents a complete ranking scheme of six chemical compounds measured in
the samples, it points out a principal diagenetic process obtained in \cite%
{Chalcraft88}. Yet this conclusion is not clear in the discriminant function
analysis.

\section*{Acknowledgment}

CYT is grateful for research support from the National Science Council,
Taiwan (ROC) and CCC is grateful for research support from both the National
Science Council (ROC) and the Institute of Geophysics (NCU, ROC).

\end{document}